%% file: coptalkpap02.tex
\documentclass[12pt]{article}
\usepackage{epsfig}

\newcommand{\BB}{{\bf{B}}}

\newcommand{\AAA}{{\bf{A}}}

\newcommand{\meanBB}{\overline{\bf{B}}}

\newcommand\ts{\times}
\newcommand\lb{\langle}
\newcommand\rb{\rangle}
\newcommand\bbB{\overline {\bf B}}

\newcommand\bbJ{\overline {\bf J}}
\newcommand\bB{\overline { B}}
\newcommand\bbA{\overline {\bf A}}

\newcommand\bfA{{\bf A}}
\newcommand\bfB{{\bf B}}
\newcommand\bfv{{\bf v}}
\newcommand\bfa{{\bf a}}
\newcommand\bfu{{\bf u}}
\newcommand\bfb{{\bf b}}
\newcommand\curl{\nabla \times}

\newcommand\emfb{\overline{\mbox{\boldmath ${\cal E}$}} {}}
\newcommand\lsim{\mathrel{\rlap{\lower4pt\hbox{\hskip1pt$\sim$}}
    \raise1pt\hbox{$<$}}}
\newcommand\gsim{\mathrel{\rlap{\lower4pt\hbox{\hskip1pt$\sim$}}
    \raise1pt\hbox{$>$}}}

\newcommand{\yjgr}[3]{ #1, {JGR,} {#2}, #3}

\newcommand{\yan}[3]{ #1, {AN,} {#2}, #3}

\newcommand{\ymn}[3]{ #1, {MNRAS,} {#2}, #3}

\newcommand{\ybook}[3]{ #1, {#2} (#3)}
\newcommand{\yproc}[5]{ #1, in {#3}, ed. #4 (#5), #2}

\input econfmacros.tex
\def\Title#1{\begin{center} {\Large {\bf #1} } \end{center}}

\begin{document}

\Title{On the Origin of Magnetic Fields in GRB Sources}

\bigskip\bigskip


\begin{raggedright}  

{\it Eric G. Blackman\index{Blackman, E.G.}\\
Dept. of Physics \& Astronomy \& Lab. for Laser Energetics, 
Univ. of Rochester\\
Rochester, NY, 14627 USA}
\bigskip\bigskip
\end{raggedright}

\begin{abstract}

Magnetic fields play a dual role in gamma-ray bursts (GRBs).
First, GRB and afterglow spectra  (the latter interpreted 
as emission from external shocks) imply synchrotron radiation in a 
magnetic field that is a significant fraction of equipartition with 
the particle energy density. Second, magnetized rotators 
with $\sim 10^{15}$ Gauss field may power GRB by transporting Poynting
flux to large distances where it dissipates and also drives an external 
shock. The field amplification at external shocks and in the engine involve 
separate processes. External shock fields are likely either seeded
by a pre-GRB wind, or are  amplified by two-stream 
plasma instabilities with MHD turbulence playing a subsequent role. 
In the engine, the large scale fields are likely produced by 
MHD helical dynamos, since flux accretion cannot easily compete with 
turbulent diffusion, and because structures must be large enough to 
rise to coronae  before diffusing.  Why helical dynamos are feasible, 
and their relation to the magnetorotational instability 
are among the points discussed.
\end{abstract}

\section{Introduction}


Magnetic fields are an intermediary between gravity and radiation. 
Often their GRB consequences are explored without addressing   
their origin. Here I focus on the latter. 

Consider GRB shocks \cite{shocks,piran00}: 
external shocks, driven by relativistic outflows acting as a piston, 
propagate into the external medium and 
accelerate particles that contribute to afterglow emission.
In contrast, internal shocks would occur inside the ejecta and
have been invoked to explain the GRB itself.  
Field amplification at the external shock 
poses the same problem whether or not the ejecta is Poynting flux 
or gas driven, but internal shocks apply only to hydrodynamic models.

Despite their advantages, Poynting flux models \cite{usov,
thompsonduncan93,reesmeszaros97} raise questions of how the emission is 
produced 
\cite{lyutikov01}, and where the  strong ordered 
field comes from. In situ helical dynamo 
amplification \cite{thompsonduncan93,moffatt78,paris01}
is well suited for amplifying
the large scale fields required for producing the 
inferred jets of GRB \cite{jets}, 
particularly in light of recent understanding of nonlinear dynamos 
\cite{paris01,brandenburg01,fb02,bf02}. 
Nonlinear helical dynamos can operate in a convectively driven star or  
in a stratified disk \cite{bnst95,bd97} unstable to the 
magnetorotational instability (MRI) \cite{balbushawley,hgb95}.

\section{Field amplification at external shocks}



Magnetic fields are likely generated as part of relativistic collisionless 
GRB shock formation. Both observations and modeling 
suggest a magnetic to  particle energy density ratio
$10^{-5}< \epsilon_m/\epsilon_{part} \simeq 10^{-1}$ (c.f. \cite{piran00}).
The ambient field of a typical surrounding ISM 
is $\sim 10^{-6}$ Gauss. Compression of the 
ambient field from a relativistic shock (of Lorentz factor $\gamma$)
would give $\sim 10^{-4}  (\gamma/10^2) (B_{ext}/10^{-6}{\rm G}) {\rm G}$, but the equipartition value is 
$B_{eq} \sim 10^2 (\gamma/10^2)(n/1{\rm cm}^{-3})^{1/2}$G,
so an amplification mechanism is needed.

The two-stream Weibel instability in the GRB context \cite{medvedev99}
may be important.
Here the field draws energy from free-streaming charged particles. 
Imagine a weak seed  magnetic field whose amplitude has a sinusoidal
perturbation in the plane perpendicular to the flow direction.
The resulting magnetic force guides charged
particles of one sign toward each other on one side of the perturbation plane
and away from each other on the opposite side of the perturbation.
This creates current channels which amplify 
the field by induction, further amplifying the currents, 
and ultimately exponentially amplifying the field. 
Initially anisotropic particle distributions become isotropic.
The magnetic fields gets the randomized energy, which  also 
helps the shock form. The instability saturates first on electrons 
and then on protons, such that the proton gyro-radius 
approaches the relativistic Debeye
length (or skin depth) $\gamma_{th}^{1/2}c/\omega_p$ 
(the scale of charge separation, where $\gamma_{th}$ is the thermalized
Lorentz factor and $\omega_p$ is the plasma frequency.). 
Saturation could produce energy equipartition between 
protons and the magnetic field.

The above saturation scale would be $\sim 8$ orders of magnitude smaller
than the source emission region.  How subsequent field growth/advection 
evolves needs more work. A few simulations \cite{nordlund02} show that 
regions of field amplification do seem to convect 
beyond the skin depth, but longer simulations are needed. 
It may actually take some distance before 
the full flow cross section participates in the instability, 
after which the field evolution is unclear.  Perhaps 
MHD instabilities (super-gyrodradius scales)
can grow and further amplify the field 
over larger scales as in supernovae \cite{jun96}.  

Another alternative: some engine models 
\cite{stellavietri98,koniglgranot} involve multiple  winds 
in the context of massive stellar collapse. The core first collapses
first to a neutron star (NS), and then later to a black hole (BH).
The GRB external shock could then propagate
into a pulsar wind ``pre-magnetized'' bubble \cite{koniglgranot}.
Outside of the light cone, the field falls off as 
$B \sim B_s \left({R_s^3\over r_{lc}^2r}\right) \sim 10^{2} \left({R\over 10^6{\rm cm}}\right)
\left({r\over 10^{15}{\rm cm}}\right)^{-1}
\left({B_s\over 10^{16}{\rm G}}\right)\left({\Omega \over 10^4{\rm s}^{-1}}\right)^2$G, where $B_s$ is the surface field, $r_{lc}$ is the light cone,
$R_s$ is the scale of the rotator and $r$ is the radius.



\section{Cold, poynting flux GRB models}

Small scale magnetic fields dissipating in the engine
could in principle drive a hot jet. But  collimation in such models
is hard to understand, and copious radiation would be
produced. The outflow then has to convert this energy to
bulk motion and then back to radiating particles. Perhaps more
appealing is the cold, baryon free, Poynting flux 
paradigm \cite{usov,thompsonduncan93,reesmeszaros97,wheeler00} 
in which particle acceleration occurs only at large distances
where flux freezing is broken (and where the acceleration is needed).
The magnetic hoop stress can provide jet collimation.

Poynting flux outflows are actually powered by rotation; the
magnetic field acts as a drive belt and determining the luminosity.
A large fraction of the rotational energy of the anchoring
compact object can in principle used (though we don't fully 
understand the dissipation mechanism yet \cite{lyutikov01}.) 
This may differ from internal shock models, where only 
the relative energy between colliding blobs in the flow is used.

Poynting flux GRB rotators can arise in non-supernova events
[(i) magnetar formed from accretion induced collapse of a white dwarf
\cite{usov,thompsonduncan93}.
(ii) merger that produces a NS/torus + BH 
\cite{reesmeszaros97}]
or in supernova events 
[(i) NS $\rightarrow$ disk + BH from fallback of material
after breakout in a partially failed or weak SN, 
with GRB powered by accretion
\cite{macfadyen99}, (ii) or powered by NS rotation 
\cite{usov,lyutikov01,ruderman00,wheeler00}
(iii) or delayed BH jet 
formation after supermassive NS forms in a  supernova (=supranova)
\cite{stellavietri98} with a 
BH accretion driven magnetic outflow propagating into an earlier pulsar wind  
\cite{koniglgranot}
(which could pre-magnetize the ambient medium for external shocks.)]
In each of these cases exponential field 
amplification (rather than linear \cite{ruderman00}) can occur, 
either in the NS  torus, in the NS itself \cite{thompsonduncan93}
or at NS boundary by MRI \cite{akiyama02} 
or helical interface dynamo as in an AGB star \cite{blackman01}
But these dynamo proposals must be evaluated in light
of recent work on nonlinear helical and nonhelical dynamos (section 5).

The basic magnetic outflow launching is either a ``spring'' 
\cite{spring} driven by toroidal field pressure in a corona
of a disk or outside the light cylinder of a magnetar,
or a centrifugal ``fling'' \cite{fling} of material along  poloidal field 
which develops a subsequent toroidal magnetic pressure.  
The maximum luminosity in either case is
$\sim B^2 r^3\Omega$ with quantities measured at the inner most 
distance $r$ at which the toroidal field becomes of order the poloidal field: 
e.g. near the surface of an accretion disk, or 
at the light cylinder for a magnetar. In the latter, the luminosity
formula becomes the dipole radiation formula when scaled to the stellar
surface.

\section{Need for in situ generation of large scale fields}

\subsection{Small scale fields shred before they escape} 

For magnetic fields to launch jets or power coronae, the field buoyancy time
must be shorter than the diffusion time in in the disk.  
The buoyancy time is $\sim h/{\overline U}_A$ where $h$ is the system thickness
scale and ${\overline U}_A$ is the Alfv\'en speed associated with the structure
of scale $L$. To escape without diffusing, 
$h/{\overline U}_A < L^2/\beta$ where the turbulent
magnetic diffusion coefficient $\beta \sim (u_2 l)$ 
with $u_2$ and $l$ the dominant turbulent speed and scale. 
(e.g. $l\sim$cube root of the turbulent cell volume $(l_x l_y l_z)^{1/3}$).
Escape then requires 
$
L^2 >  l h (u_2/{\overline U}_A)
$ 
The ratio on the right is typically $\gsim 1$ (more on that later)
so $L^2 > l h$ is required for escape.


Now consider a Shakura-Sunayev \cite{shakura73} disk viscosity 
$\nu_{ss}\sim \beta= u_{2} l \sim \alpha_{ss} 
c_s h \sim u_{2}^2 /\Omega \sim u_{A,2}^2/ \Omega,$
where $u_{A,2}$ is the Alfv\'en speed associated with the 
turbulent $B$-field and  $u_{2}\sim u_{A,2}$ follows
from simulations \cite{hgb95,bnst95}. 
Then $\alpha_{ss} \sim u_{A,2}^2/c_s^2$
or $u_{A,T} \sim \alpha_{ss}^{1/2} c_s$,
so that $l =\alpha_{ss}^{1/2}h$.
The escape condition $L^2>lh$ then  gives 
$L >  \alpha_{ss}^{1/4} h $
{\it smallest} dimension of the {\it largest} structures 
since the former determines the diffusion time.
The escape condition is not satisfied in the nonhelical
MRI dynamo (see maximum $k_y,k_z$ in fig. 4 of \cite{hgb95}).
Note also if $L=l\sim\alpha^{1/2}H$, then escape requires 
$\alpha_{ss}\sim 1$.


\subsection{Accretion of flux can't easily beat turbulent diffusion}

Can the large scale fields required by jets be accreted onto a central 
engine? Though 3-D MHD simulations are needed, 
the answer  seems to be ``no'' particularly for thin disks.
The reason is that MHD the turbulence produces 
a turbulent  magnetic diffusivity $\beta\sim \nu_{ss}$. 
To determine if a vertical field can be accreted in an MHD turbulent disk,
the ratio of rates of flux advection to diffusion    
is needed.  The dominant variation of the field strength
is likely in the vertical direction across the thickness of the disk and 
the dominant variation of the radial velocity is in the $r$ direction.
The above mentioned ratio is then 
$\nabla\times({\overline {\bf v}}\times \bbB)/(\beta \nabla^2 \bbB)
\sim (h^2 / R^2)(\alpha_{ss} h c_s / \beta)
\sim {h^2/ R^2}$. The diffusion term thus dominates for a thin disk.
This was shown with a numerically 2-D calculation in \cite{lubow94}.
and motivates an in situ helical dynamo which can overcome diffusion.

\section{MHD in situ amplification}

It is useful to categorize dynamos into two classes: direct and inverse.
Direct dynamos sustain magnetic fields on scale at or below 
the dominant turbulent scale.
They do not require helical turbulence although helicity 
does influence the resulting magnetic spectrum. 
In contrast, an inverse dynamo describes  amplification 
on scales larger than that of the dominant scale of the turbulence.
The label ``inverse'' is used to suggest an inverse
cascade. The turbulence must be  helical 
to generate and sustain large scale flux over times longer than an
eddy turnover time in order to drive and collimate jets.

\subsection{Direct dynamo}

In a nonhelical direct dynamo, the field grows by line stretching in a 
turbulent random walk. The turbulence can be 
externally driven by isotropic forcing, or self-generated  
by an angular velocity gradient. In either case, 
turbulent stretching compensates for exponential
decay from turbulent diffusion. The latter operates
because the random motions of the gas also mix the field lines,
inducing a cascade to small scales where dissipation occurs. 
In 3-D, a steady state balance can be achieved.
In the saturated state, 
the magnetic energy $\sim$ turbulent kinetic energy 
below the dominant turbulent scale, but the spectral shape is a subtle issue.

\subsubsection{Forced turbulence}

In numerical experiments where turbulence is forced isotropically 
in a periodic box without helical correlations, 
the field piles up at the resistive scale \cite{kida91,maroncowley}
for isotropic viscosity and for magnetic Prandtl number $\equiv \nu/\lambda>1$
(where $\nu$ is the microphysical viscosity and $\lambda$ is the magnetic
diffusivity).  When the turbulence is forced with a
a fractional kinetic 
helicity $f_h\sim \lb\bfu\cdot\curl \bfu\rb/k_2 u^2 >0.4$ ($k_2$ is the 
forcing wavenumber $\sim 5$ in Refs. 
\cite{brandenburg01,maronblackman}), the direct dynamo
then piles up magnetic energy at the forcing
scale of the turbulence, not the resistive scale \cite{maronblackman}.
(A second peak also develops at larger scales due to the inverse
dynamo discussed in section 5.)


For $f_h=0$, $Pr>1$, and isotropic viscosity,  
a nonlocal effect induces the field  pile up on small scales 
\cite{maroncowley,scmm}:
forcing scale motions directly stretch the field into 
long magnetic folds which can't unwind for  $Pr_M > 1$ 

\bigskip


\subsubsection{Nonstratified MRI: direct nonhelical dynamo, no sustained flux}

In a disk (or star) with a weak magnetic field, 
the MRI is driven by the angular velocity gradient 
and is thus different from isotropically forced turbulence. 
In the absence of stratification, and thus the absence of 
a pseudoscalar helicity, the MRI does not produce magnetic fields on scales 
larger than the scale of the developed turbulence.  The largest 
scale of the field in the toroidal direction does 
approach the scale of the box height in 
simulations \cite{hgb95}, 
but the kinetic turbulence also approaches this scale.  
The marginally unstable (and dominant) growth
mode of the MRI satisfies $k u_A \sim \Omega$ \cite{balbushawley}
so that as the magnetic field is amplified from the
instability, larger and larger scales
become unstable to growth. The largest scale
that can grow is that associated the scale height $h$.
Without stratification or helicity, 
one would expect the largest scale of the turbulence
to be of order the largest scale of the the field and this is 
what is seen (Fig. 4 of \cite{hgb95} where $k_y$ is the toroidal 
direction.) 

Note that the MRI, unlike the isotropic forced 
turbulence, produces a nearly Kolmogorov spectrum
in the field even without any helicity. The role of large
scale shear (rather than helicity) in keeping the field from
piling up on the resistive scale needs to be further investigated.
That being said, note that the flux associated with the field
on the largest scale in the nonstratified MRI would 
change every turnover time as the field is shredded.   
The field is large scale in the nonstratified MRI only in the sense that 
the box height limits the scale of the turbulence NOT because
there is any sustained ordered flux on scales larger than the turbulence.  
Furthermore, when the box 
is periodic, the net flux cannot grow, as it is conserved. 
Note also that the scale of the largest field structure in the vertical 
and the radial $(k_x)$ directions are smaller than in the toroidal direction.
This is important in the context of section 4.1.
Because of the rapid change in sign of the flux and 
the constraints of section 4.1, a nonstratified nonhelical
MRI case would {\it not} produce the fields required for jets---
but real rotators do have stratification and a helical
version would work (see sect. 5.2.3).


\subsection{Helical inverse dynamo}

The helical inverse dynamo amplifies field on scales larger than that of
the turbulence.  This dynamo is the most plausible
for the production of large scale fields for cold Poynting flux GRB models.

Here is the basic mechanism  \cite{moffatt78,zeldovich83}: 
Consider an initially weak toroidal (=encircling the rotation axis)
loop of the magnetic field embedded in the rotator.
With an outwardly decreasing density gradient 
a rising turbulent eddy threaded by a magnetic field   
will twist oppositely to the underlying global rotation 
to conserve angular momentum.
Statistically, northern (southern) eddies  
twist the field clockwise (counterclockwise). This is the ``$\alpha_d$'' 
effect and the result is a large scale poloidal field loop.
Differential rotation shears this 
loop (the ``$\Omega$''-effect). The bottom part 
reinforces the initial toroidal field and the top part diffuses
away. The result is exponential growth.
Inside a star the $\alpha_d$ effect can be supplied by
neutrino driven convection + a vertical density gradient
For an MRI driven system, such stratification is also required.








This is revealed mathematically by averaging the magnetic induction 
equation over a local volume and breaking all quantities
(velocity $\bf U$, magnetic field $\bf B$ in Alfv\'en velocity
units, and  normalized current density ${\bf J}\equiv {\curl {\bf B}}$)
into their mean (indicated by an overbar) and 
fluctuating (lower case) components. The result is 
\cite{moffatt78}:
\begin{equation}
\partial_t\meanBB= \curl(\alpha_d\meanBB+{\overline {\bf U}}\ts \meanBB) 
+(\beta+\lambda) \nabla^2\meanBB.
\end{equation}
%
The ${\overline {\bf U}}$ 
term incorporates the $\Omega$-effect,
the $\beta$ term incorporates the turbulent diffusion 
(assume constant $\beta$) 
and the first term on the right incorporates the $\alpha_d$-effect.
The quantity $\alpha_d$ is given by \cite{bf02,pfl} 
$\alpha_d=\alpha_{d0}+ (\tau/3){\overline {\bfb\cdot{\curl \bfb}}}$,
where $\alpha_{d0}= -(\tau/3)
{\overline {\bfu\cdot\curl\bfu}}$, $\tau$ is a 
turbulent damping time and ${\overline {\bfu\cdot\curl\bfu}}$ is the kinetic
helicity. 
In textbooks, $\alpha_d=\alpha_{d0}$, but properly 
including the magnetic forces that backreact
on the velocity driving the field growth lead to
the above correction term. This  
is studied formally 
in Ref. \cite{bf02}.

%


From the form of $\alpha_d$, it is evident that the correction
term can offset $\alpha_{d0}$ and quench the dynamo
\cite{fb02,bb02,bf02}.  
The principle of magnetic helicity conservation determines
the extent to which $\alpha_d$ is suppressed.
The magnetic helicity, a volume integral
$H\equiv \int\AAA\cdot\BB d V\equiv\lb\AAA\cdot\BB\rb V$ 
satisfies \cite{woltjer58,berger84}
\begin{equation}
\partial_t H=
-2\lambda C-\mbox{surface terms},
\label{magcons}
\end{equation}
where $\AAA$ is defined by  $\BB=\nabla\times\AAA$, 
and the current helicity $C$ is defined by 
$C\equiv\lb{\bf J\cdot\bfB}\rb V$.
Without non-diffusive surface terms 
$H$ is well conserved:
for $\lambda\rightarrow 0$ the $\lambda$ term in (\ref{magcons}) 
converges to zero \cite{berger84a}.
Since $H$ is  a measure of ``linkage'' and ``twist'' of field 
lines \cite{berger84}, its conservation implies that the $\alpha_d$-effect does not
produce a net magnetic twist, just positive and negative magnetic 
twists on different scales.
\cite{fb02,seehafer96}.

The importance of the scale segregation 
is most easily seen in a  two-scale approach.
Write $H=H_1 + H_2$, where 
$H_1=\lb\bbA\cdot\bbB\rb V$
and $H_2=\lb{\overline {\bfa\cdot\bfb}}\rb V$ correspond to 
volume integrated large and small scale contributions respectively.
Also, $C=\lb\bbJ\cdot\bbB\rb V+\lb{\overline{{\bf j}\cdot\bfb}}\rb V
=k_1^2H_1+k_2^2H_2$,
where $k_1$ and $k_2$ represent the wavenumbers (inverse gradients) 
associated with the large and small scales respectively,
and the second equality follows for a closed system.

Now relate $\bbB$  to $H_1$.
Define $\epsilon_1$ such that the large scale magnetic energy
$\lb\bbB^2\rb V= H_1{k_1/\epsilon_1}$ 
and where
${0< |\epsilon_1| \leq 1}$, where $|\epsilon_1|=1$
only for a force-free helical large scale field 
(i.e. for which $\bbJ || \bbB$ so that  the force $\bbJ \ts \bbB =0$).
In a northern hemisphere $\epsilon_1 > 0$.
The conservation equations analogous to (\ref{magcons}) 
for $H_1$ and $H_2$ are 
\begin{equation}
\partial_t H_1 = 2(\lb\alpha\rb {k_1/\epsilon_1} 
-\lb\beta \rb{k}^2_1)H_1-2 \lambda k^2_1 H_1
-\mbox{surface terms} 
\label{h1}
\end{equation}
and 
\begin{equation}
\partial_t H_2=-2( \lb\alpha \rb
{k_1/\epsilon_1} -\lb\beta\rb k^2_1)H_1 
-2\lambda {k^2_2 H_2}
-\mbox{surface terms}
\label{h2}
\end{equation}
where 
$\lb\alpha\rb=
(\lb\alpha_{0}\rb + {1\over 3}\tau k^2_2
H_2/V)$.
The case without non-diffusive surface terms and with $\epsilon_1=1$ 
 is called an $\alpha_d^2$ dynamo.
The solution \cite{fb02,bf02,bb02}. 
shows that for initially  small $H_2$ but large $\alpha_{0}$,
$H_1$ grows. Growth of $H_1$ implies the oppositely signed growth of $H_2$.
This $H_2$ backreacts on $\alpha_{0}$, 
ultimately quenching  $\lb\alpha_d\rb$ and the dynamo. The theory 
quantitatively matches simulations of \cite{brandenburg01}.
At early times, kinematic growth is 
unimpeded, and the large scale field can grow to 
${\overline U}_A^2 \sim f_h (k_1/k_2)u_2^2$, where $f_h$ is the fractional
kinetic helicity. (In general $f_h \le 1$ must be estimated,
but  $f_h\sim 1$ will be later derived for young NS/torus engines of  GRB.)
Eventually, the  small-scale magnetic helicity backreacts on the kinetic
helicity, suppressing the growth rate to a resistively limited
value. Ultimate saturation occurs at ${\overline U}_A^2 
\simeq f_h (k_2/k_1)u_2^2$ but growth beyond the kinematic regime is 
too slow to be of relevance for GRBs.



\subsubsection{Role of the Rossby Number $R_o$}

When differential rotation operates over the same scale
as $\alpha_d$, a rotation  to eddy turnover time ratio  
$R_o\equiv t_{rot} /t_{ed}=v_2k_2/\Omega > 1$ favors
an $\alpha_d^2$ dynamo over an $\alpha_d-\Omega$ dynamo.
This follows from comparing the $\Omega$ and $\alpha_d$ 
terms effects in the $\partial_t {\overline B}_\phi$ equation. But
the $\Omega$ layer could be $<<$ than the $\alpha_d$
layer \cite{blackman01}. Then $R_o$ is not the determining parameter,
but either dynamo produces the desired large scale fields. 
A more important role of $R_o$: for the 
growth term in (\ref{h1}) to overcome decay,
$|\epsilon_1|\beta k_1/\alpha_{d0}\sim |\epsilon_1| R_o < 1$, using 
$k_1h \sim 1$, $\beta \sim u_2/k_2$ and using 
$\alpha_{d0}\sim \Omega \cdot \nabla\rho/\rho k_2^2\sim \Omega/h k_2^2$ 
\cite{moffatt78,zeldovich83} for the case $R_o \gsim 1$.
Since $|\epsilon_1| \le 1$ the condition  can be met.


\subsubsection{Application to young NS/torus GRB progenitors}

Here I cavalierly apply the nonlinear results of sect. 5.2 to 
young NS-like MHD rotators typical of Poynting flux GRB engine models
\cite{thompsonduncan93}. Without buoyancy, the amplification term 
in the kinematic regime is 
$\sim \alpha_{d0} k_1-\beta k_1^2$.  Although buoyancy 
dominates the $\beta$ term when
$1/k_1{\overline U}_A < 1/\beta k_1^2$ (where $k_1$ is used  
for the inverse system size and the wavenumber of $\overline B$)
for $f_h\sim 1$, it does not limit 
${\overline U}_A$ calculated in sect. 5.2 because a buoyancy
time $H/{\overline U}_A$ equals a growth time when 
$f_h v_2 \sim {\overline U}_{A}$, which is $(k_2/k_1)^{1/2}$ larger than the
saturation value from mag. helicity conservation for $f_h=1$.

The $f_h$ for the young 
NS case can be estimated  from $f_h\sim \alpha_{d0}/u_2$ 
taking  $u_2\sim \ts 10^8 F_{39}$cm/s
as the neutrino driven turbulent convection velocity, 
and $F_{39}$ is the neutrino heat flux in units 
$10^{39}$ erg/sec cm$^{-2}$ \cite{thompsonduncan93}.  
Using  $\alpha_{d0}$ from Ref. \cite{moffatt78} 
and the  NS numbers  \cite{thompsonduncan93}, 
$\alpha_d\sim 10^{8}\left({\Omega\over 6 \ts 10^3 {\rm sec^{-1}}}\right) \left({l\over 2\ts 10^5 {\rm cm}}\right)^2 \left({h\over 10^{6}{\rm cm}}\right)^{-1}$ cm/s, 
so $f_h\sim 1$. From section 5.2 the maximum field of a closed system 
at the end of the kinematic regime is then 
${\overline U}_A =  u_2(k_1/k_2)^{1/2}$. 
As applied to the young NS convection zone,
$u_2$ corresponds to the convective equipartition field $B_{eq}$ for $f_h\sim 1$, 
and for $k_1/k_2 \sim 1/5$ we then have 
${\overline B}\sim (k_1/k_2)^{1/2}B_{eq}\sim 4\ts 10^{15}F_{39}$G,
with growth time is $\sim 10^{-3}$ sec. This is sufficient for GRB.
This estimate from an $\alpha_d^2$ dynamo 
is a lower limit to that from an   
$\alpha_d-\Omega$ dynamo since the latter gains from  
shear, such that $|\epsilon_1| < 1$ \cite{bb02}.  
But since $f_h\sim 1$ corresponds to $|\epsilon_1|R_o \sim k_1/k_2$,
for $R_o\sim 1$, the lower limit $\sim$ actual value.






\subsubsection{Helical MRI dynamo} 

In a sheared rotator, the presence or absence
of an MRI dynamo {\it does not} depend on helicity
but the additional presence of an {inverse helical dynamo} 
(described in section 5.2) does.  Only a helical dynamo can 
sustain a large scale ordered magnetic field and magnetic flux 
over long time scales. This has been demonstrated in
stratified shearing box simulations \cite{bd97}.
There MRI turbulence can generate its own kinetic or current helicity 
producing a finite $\alpha_d$.
A global large scale flux is also produced if periodic
boundary conditions are avoided.
Figure 2 of \cite{bd97} clearly shows the effect of a helical
dynamo operating in an MRI simulation, where a large scale
field with cycle period of 30 orbit periods is seen.
Thus the large scale fields that satisfy the escape condition of section 4.1
can in general also be produced 
by the MRI, but only when density stratification is included.
Using the MRI in spherical geometry \cite{balbusstar} 
to generate field at the boundary of the NS GRB progenitor \cite{akiyama02} 
should involve the MRI as a potential source (in addition to convection) 
of a helical dynamo, not just a nonhelical one.

\section{\bf Summary points}

(1) The problem of field amplification in  external shocks is the same
for both hot and Poynting flux engine models.
Weibel instability + MHD turbulence may work in succession.
(2) Large scale fields in the engine are required for GRB
cold Poynting flux jets and are difficult to accrete because of turbulent
diffusion. At minimum, flux accretion  requires very thick $h\sim R$ disks.
(3) Buoyancy must beat turbulent diffusion to get field in coronae to 
produce jets. Even this requires large scale fields. 
(4) A balance between 
helical dynamos and buoyancy can produce the required GRB engine fields.
The expected large scale field strength is 
proportional to the ratio of turbulent  to system scale for  
fully nonlinear helical dynamos. Strong large scale fields result.
(6) The magnetorotational instability (MRI) can lead to 
a nonhelical or helical dynamo, but only the helical
version (expected in nature's stratified, sheared rotators) 
favorably produces the large scale fields required for jets.


Thanks to B. Paczynski and V. Pariev for recent discussions,
and to the Dept. of Astrophysical Sciences, Princeton Univ. for hospitality.

\def\Discussion{
\setlength{\parskip}{0.3cm}\setlength{\parindent}{0.0cm}
          {\Large {\bf Discussion}} }
\def\speaker#1{{\bf #1:}\ }
\def\endDiscussion{}

\Discussion

blackman@pas.rochester.edu
Speaker: Eric Blackman

\speaker{A. Hujeirat, Max Planck Institute}
Do you agree that BH-instability is a transient phenomena 
because the field in sims. never reach $u_A=c_s$?  Even if it did,  
the field will flow up through Parker instability.  If you switch on 
dynamo of Ref. \cite{tout} can you produce large scale magnetic fields 
appropriate for jets? 

\speaker{Blackman}
(1) In the saturated (steady, not transient) state, 
MRI simulations usually show $u_A < c_s$.
Since $u_A/c_s\sim \alpha_{ss}^{1/2}$ (see section 4.1) the 
question of $u_A$ is a question of predicting  $\alpha_{ss}$
which is not yet understood, though $\alpha_{ss} \sim 1/2\pi$ is not
yet theoretically ruled out (see below, Shibata question).
Simulations do however show  $u_A\sim u_2$ (turbulent equipartition). 
(2) The dynamo of \cite{tout} might be important (see constraint of sec.
4.1) but in a helical generalization: they discuss the end
state of an inverse cascade,  not the time evolution.
A helical dynamo is needed to make this happen fast enough for GRB.


\speaker{A. Brandenburg, NORDITA}
A crucial question is where the kinetic helicity comes
from. You mentioned thermal buoyancy but simulations haven't 
yet shown this. Instead of expansion of rising tubes we see
inflow along field lines toward the field maximum, i.e.
contraction and hence the opposite sign of the dynamo $\alpha_d$.

\speaker{Blackman}
This is indeed important, but 
fortunately the Poynting flux doesn't care about the sign of $\alpha_d$
as long as its magnitude is not too small. Also, since  
$\alpha_d\propto \overline{\bfu\cdot\nabla\times \bfu} - \overline{\bfb\cdot\nabla\times \bfb}$  the current helicity could actually be 
the main  driver in an MRI unstable disk.

\speaker{K. Shibata, Kyoto Univ.}
What is the relation between your theory and 
the mechanism of MRI field saturation?
How do you think this issue will be resolved?

\speaker{Blackman}
As shown in section 4.1, 
asking how large $u_A/c_s$ gets is also asking  how large $\alpha_{ss}$ 
should be.  It is not yet clear why  $\alpha_{ss}$ should not be 
$\sim (1/2\pi)^{1/2}$;
one might say that the field produced
either by helical or nonhelical dynamos is limited in the 
thin disk MRI framework only by the strength that keeps the disk thin, 
or by the strength which for which the vertical Alfv\'en crossing
time equals the orbital period. Since the wavenumber of maximum
growth of the MRI satisfies $ku_A \sim \Omega$, as the field energy
grows (i.e. as $u_A$ grows), $k$ decreases, but not smaller than
that associated with the disc thickness. This would imply
 $\alpha_{ss}^{1/2} \sim u_A/c_s\sim 1/2\pi$.
This argument does not invoke anything about  helical large scale 
dynamo theory except that the 
ratio could be lowered in practice by large structures as they
remove field from a thin disk. Sometimes small boxes 
show values of $\alpha_{ss}<< (1/2pi)^{1/2}$  
but numerical simulations have not converged on a universal value.  
In thick disks we might expect a higher value than for thin disks, 
and there are no global simulations of thin disks yet. 
More theory and simulation are needed to pin this down. It is 
not even clear that $\alpha_{ss}$ is a constant; it could depend on radius.

\speaker{S. Kulkarni, Caltech}
Focusing on the afterglow external shock, on 
what time scales can fields grow?  I draw your attention to
fields in young supernovae (weeks to months old).
The inferred $B$ strengths are far in excess of ISM fields.
Are these generated by the mechanism(s) you discussed for the 
external shock or are they compressed fields in the wind of the progenitor ?

\speaker{Blackman}
The plasma (two-stream Weibel) instability  would saturate
on incredibly short time scales, but the associated fields generated are
initially on too small a scale  without further hydrodynamic evolution.
I think the shocks eventually become unstable to MHD turbulence. 
This turbulence can then further amplify the fields to 
near equipartition with turbulent motions on a dynamical time
scale of the eddies (by the ``direct'' nonhelical type dynamo discussed)
which should be of order a dynamical shock 
crossing time. Ref. \cite{jun96} showed that this is promising;
fields were amplified to within a factor of 10 of the observed
fields. Additional simulations are needed.

\endDiscussion

\bigskip

\vfill
\eject

\end{document}

%% file: econfmacros.tex
\def\beq{\begin{equation}}
\def\eeq#1{\label{#1}\end{equation}}
\def\eeqn{\end{equation}}

\def\beqa{\begin{eqnarray}}
\def\eeqa#1{\label{#1}\end{eqnarray}}
\def\eeqan{\end{eqnarray}}

\let\bar=\overbar

\def\Dslash{\not{\hbox{\kern-4pt $D$}}}
\def\dslash{\not{\hbox{\kern-2pt $\del$}}}

\def\msb{{\bar{\ssstyle M \kern -1pt S}}}

